# Amorphous Alloys, Degradation Performance of Azo Dyes: Review

Hasan Eskalen[1], Celal Kursun[2*], Mikail Aslan[3], Mustafa Cesme[4] and Musa Gogebakan[2]

[1]*Department of Material Science and Engineering, Institute of Science, Kahramanmaras Sutcu Imam University, Kahramanmaras, 46100, Turkey*

[2]*Department of Physics, Faculty of Art and Sciences, Kahramanmaras Sutcu Imam University, Kahramanmaras, 46100, Turkey*

[3]*Department of Metallurgical and Materials Engineering, Gaziantep University, Gaziantep, Turkey*

[4]*Department of Chemistry, Faculty of Art and Sciences, Kahramanmaras Sutcu Imam University, Kahramanmaras, 46100, Turkey*

## Abstract

Today freshwater is more important than ever before and it is contaminated from textile industry. Removal of dyes from effluent of textile using amorphous alloys has been studied extensively by many researchers. In this review article it is presented up to date development on the azo dye degradation performance of amorphous alloys, a new class of catalytic materials. Numerous amorphous alloys have been developed for increasing higher degradation efficiency in comparison to conventional ones for the removal of azo dyes in wastewater. One of the objectives of this review article is to organize the scattered available information on various aspects on a wide range of potentially effective in the removal of dyes by using amorphous alloys. This study comprises the affective removal factors of azo dye such as solution pH, initial dye concentration, and adsorbent dosage. It was concluded that Fe, Mg, Co, Al and Mn-based amorphous alloys with wide availability have appreciable for removing several types of azo dyes from wastewater. Concerning amorphous alloys for future research, some suggestions are proposed and conclusions have been drawn.

[*]Corresponding author:

Department of Physics, Faculty of Art and Science, Kahramanmaras Sutcu Imam University, Kahramanmaras, 46100, Turkey,

Tel.:+ 90 344 300 2540; Fax: +90 344 300 1352
E-mail address: celalkursun@ksu.edu.tr or celalkursun@hotmail.com

**Contents**

# 1. Introduction

## *1.1. The status of freshwater resources*

Although about 71 % of earth surface are covered by water, only very small amount of it is vital for humans since water sources consist of nearly 97.5 % salty water and only 2.5 % freshwater. Small amount of the freshwater are accessible for human usage. The human could use merely 0.007 % of all water (Petersen et al., 2017). Fig. 1 shows the water distribution on the world mentioned above. At the beginning of twenty-first century about one out of five of humans are suffering from freshwater shortage and this will be much worse at the half of this century according to the report of UNESCO (Deb and Dutta, 2017). In addition, although approximately 780 million people do not still have access to proper resources of drinking water (Qu et al., 2013), the limited freshwater sources have been contaminated due to agricultural, industrial and domestic activities (Cosgrove and Rijsberman, 2014; Matheyarasu et al., 2015; Petersen et al., 2017).

Thus, water contamination has been an important issue for various fields of researchers and engineers to be eliminated. Until 2016, the number of scientific papers titled "wastewater" can be seen in Fig. 2. Although there are only 4 publications in 1992, over 3,500 papers have been published in 2016. On the other hand, only about 10 % of wastewater is treated (Matheyarasu et al., 2015) and commonly used in wastewater treatment processes while the remaining part of wastewater cannot be entirely cleaned from many toxic materials which are basically azo dyes and heavy metals (Luo et al., 2014b; Sadegh et al., 2017; Saha et al., 2017).

## *1.2. Effect of azo dyes on formation of wastewater*

Dyes are simply divided into two main groups which are organic and synthetic. The most common use of synthetic dyes is azo dyes with above 50 % of usage of all dyes (Ajmal et al., 2014). Today, more than 100.000 dyes with different structures have been synthesized (Sekuljica et al., 2015). The global market production of synthetic dyes exceed 10 million metric tons and it is continue to grow in the rate of 3-4 % (Ahmet Gürses et al, 2016), which implies the total production of azo dyes are more than 5 million metric tons. The most consumption of azo dyes that leads to water contamination is due to the textile industry since it generates large volume of effluent (Chequer et al., 2013). To dye 1 kg of cotton, 30-60 g of dyestuff and 70-150 L water are required (Sen et al., 2016; Wang et al., 2014) and it is reported that up to 50 percent of used dyes directly or indirectly contaminated water sources (Muhd Julkapli et al., 2014; Ramya et al., 2017; Sekuljica et al., 2015; Sen et al., 2016). Wastewater from dyes not only cause environmental pollution (Amir et al., 2016; Asgher and Bhatti, 2012; Lim et al., 2013; Sarkar et al., 2014; Zaini et al., 2014) and contamination of freshwater source and land, but also carcinogenic effect (Chahbane et al., 2007; de Jong et al., 2016; Fonovich, 2013; Gupta et al., 1990; Sharma et al., 2003; Soriano et al., 2014; Zeng et al., 2014; Zhou et al., 2016).

## 2. Current treatment methods and materials for dye removal

Generally, methods of dye wastewater treatment are divided into three categories that are physical, chemical and biological methods. Adsorption, ion exchange, chemical precipitation, coagulation-flocculation, oxidation, electrochemical treatment, membrane filtration biological treatment and combined techniques are most used dye removal strategies (Ahmad et al., 2015) of dye removal methods. Advantages and disadvantages of some techniques of dye removal industrial wastewater are given in Table 1 (Crini, 2006; Garg et al., 2004; Robinson et al., 2001; Yagub et al., 2014). Traditional treatment methods for azo dye removal have been reviewed by several authors (Bethi et al., 2016; Brillas and Martínez-Huitle, 2015; Deb and Dutta, 2017; Forgacs et al., 2004; Fu et al., 2014a; Gupta et al., 2009; Holkar et al., 2016; Pearce et al., 2003; Sadegh et al., 2017; Salleh et al., 2011; Sen et al., 2016; Singh et al., 2015; Yagub et al., 2014). Activated carbon (Malik, 2004; Namasivayam and Kavitha, 2002; Santhy and Selvapathy, 2006), nanomaterials and nanocomposites (Alkaim et al., 2015; Chawla et al., 2017; Ozmen et al., 2010; Prashanth et al., 2017; Selen et al., 2016; Sohni et al., 2017), agricultural solid waste (Khatoon and Rai, 2016; Namasivayam and Kavitha, 2002; Singh et al., 2017), clay (Ehsan et al., 2017; Hadjltaief et al., 2016; Ho et al., 2001; Ngulube et al., 2017), peat (Ho and McKay, 1998; Zehra et al., 2016; Zehra et al., 2015) , egg shells (Abdel-Khalek et al., 2017; Tsai et al., 2008), tea waste (Wen et al., 2017), polymers (Crini, 2003; Tu et al., 2017; Yao et al., 2016), wood (Ferrero, 2007; Leechart et al., 2009; Poots et al., 1978) are some of the used materials at azo dye wastewater treatment but these materials can not sufficiently reduce azo dye and their environmental impacts (Wang et al., 2014b; Xie et al., 2016). Zero-valent metals have been used for new azo dye wastewater treatment materials and among them zero-valent iron exhibit superior features at high efficient dye removal, low cost production and simple operation (Fu et al., 2014b; Ruiz et al., 2000; Xie et al., 2016). However, easily oxidization and rapidly decaying of degradation efficiency are two main problems of these materials (Xie et al., 2016). Because of this deficient, the new functional materials should be studied. Amorphous alloys could be new alternative materials with superior properties compared to zero-valent metals. To the best of our knowledge amorphous alloys in the field of dye wastewater have been never reviewed previously.

## 3. Properties and Production Methods of Amorphous Alloys

The discovery of amorphous alloys which are known as one of the metastable materials has marked a new era in atomic structure of solids and crystallography. The searches reveal that these materials may exhibit unique structure and outstanding properties. The atomic packing of amorphous materials in contrast to crystalline alloys is not ordered and indicates a random distribution (Inoue, 1999; Inoue, 2000).

*3.1. Properties of amorphous alloys*

The amorphous alloys have received much attention because of their remarkable properties in comparison to crystalline alloys. They have good corrosion and oxidation resistance, low electrical conductivity, low thermal conductivity, good magnetic properties, ultrahigh strength and high hardness which are certain of significant mechanical properties at room temperature (Gebert et al., 1999; Janik-Czachor et al., 2002; Kilian and Schultz, 1988; Kim et al., 2005; Kim et al., 2004; Yang et al., 2007). Due to these excellent properties, amorphous alloys have been extensively studied from structural and functional applications since the first synthesis by Duwez et al. They succeed production of Au-Si binary amorphous alloy by rapid solidification technique in 1960 (Klement et al., 1960) and since then many binary, ternary, quaternary and quinary amorphous alloys have been manufactured. As well as above properties of the amorphous alloys, a new characteristic of them can be added. It is good degradation performance of azo dyes which contain carcinogenesis. It was schematized the amorphous alloys degrade to azo dyes in Fig.3. As shown in Fig. 3, the azo dyes has orange colour at initial. After the amorphous materials mix into the azo dyes, the orange colour transform into transparent one due to degradation of azo dyes. Therefore, these high quality materials have provided cleaner water to environment without any carcinogenesis.

*3.2. Production methods of amorphous alloys*

The amorphous materials can be synthesized by different techniques. Some of these techniques are rapid or conventional solidification, mechanical alloying and suction casting. In the rapid solidification technique, the amorphous alloys are manufactured in ribbon form. To produce the ribbons is required to high cooling. Thus the amorphous materials are produced by rapid solidification of a liquid. In this technique, cooling rates should be $10^5$ to $10^9$ K/s to avoid the nucleation of high-temperature equilibrium phases. This technique is schematized in Fig.4. As shown in Fig. 4, the melt are ejected on the copper wheel by Argon (Ar) gas pressure. Then the amorphous material is obtained in ribbon form. In the mechanical alloying technique, the materials are manufactured in powder forms, which can be compacted in desired shapes and dimensions for practical application. One of the most significant advantages of mechanical alloying technique can be controlled the structures of produced amorphous alloy during proceeding process. Therefore, it is very easy to obtain desired microstructures which are so difficult or impossible to produce with other techniques. The working principle of mechanical alloying is schematized in Fig. 5. As shown in Fig.5, the powders in the stainless steel cups remain between balls which collide with each other and the amorphous material are produced. In the suction casting technique, an arc-melted material is sucked into a copper mould, which can be chosen different length and diameter. The sucking process is done by negative pressure in the mould relative to the main chamber. Cooling rate of this technique rely on certain parameters which are mould temperature, casting temperature, interfacial heat transfer and mould geometry (Laws et al., 2009). The furnace where the materials are melted in arc melting machine is schematized in Fig. 6.

## 4. Degradation Performance of Amorphous Alloys

The most widely investigated amorphous alloys for azo dye degradation from wastewater are Fe-based, Mg-based and some miscellaneous alloys. In this review, amorphous materials are examined at three different categories. The most studied alloys are Fe-based alloys since their degradation efficiency compared with zero-valent irons. Although there are studies just only use metallic glass as a catalytic material, hydrogen peroxide, persulfate and solar light lamp are used with some combined techniques.

### *4.1 Fe-based amorphous Alloys*

Fe-based amorphous alloys which are known as a kind of zero-valent iron material have characteristics of a metastable structure and chemical homogeneity in particular. This makes Fe-based amorphous alloy more corrosive resistance and more chemically reactive on the whole part of the surface instead of the defects such as the grain boundary in crystalline materials (Tang et al., 2015). Due to these merits, it demonstrates that Fe-based amorphous alloys are higher degradation efficiency compared to the conventional techniques (Das et al., 2015; Wang et al., 2012a; Yang et al., 2013). The removal of pollutants including azo dyes in water by means of a reduction reaction using Fe-based amorphous alloys has been developed recently. As an alternative catalysis for the crystal typed zero-valent iron, Fe based amorphous alloys has been studied (Deng et al., 2017; Xie et al., 2016) since crystalline type zero-valent iron is easily oxidized (rust) and its degradation efficiency decays rapidly even if the most commonly used metal is crystalline type zero-valent iron, which shows low cost, simple operation, and high efficiency. In addition, Fe-based amorphous alloys offer a lower cost alternative, and thus more attractive for commercialization. Azo dye degradation studies by using iron based metallic glasses are summarized and presented in Table 2.

Fe–Si–B amorphous ribbon was successfully fabricated using a melt spinning method by Wang et.al (Wang et al., 2014). In related study, Rhodamine B, which is a kind of azo dyes, could be degraded by the catalyst with low dosage of $H_2O_2$ (see Table 2). In addition, the catalyst demonstrated good stability and reusability. Another study was conducted by Wang et.al (Wang et al., 2012). As a result of this study, the powders they produced could completely remove the Direct Blue 6 azo dye ($C_{32}H_{20}N_6Na_4O_{14}S_4$ ) from aqueous solution in short time as shown in Table 2.

Although Fe-based metallic glasses have been found more effective for degrading azo dyes, many issues are still inadequately understood and have need of further investigation which consists of how it is influenced by the degradation environment and by the atomic and electronic configurations of the amorphous structure, how to improve the optimized Fe-based amorphous alloy systems with higher degradation efficiency and service life and finally solving the underlying mechanism of the degradation process.

*4.2 Mg-based Amorphous Alloys*

Apart from the ferrous alloys, for degrading the water contaminants, Mg-based amorphous alloys are preferred due to the resistant to rusting, cheap material to obtain, widely tunable compositions and intrinsic brittleness that provides glass forming ability. In addition, during the degradation processes, Mg atoms not only offer high surface area for the adsorption, but also improve the electron transfer between metal atoms and organic molecules. Thus, the combination of these properties makes Mg-based amorphous alloys show high degradation efficiencies of azo dyes than the corresponding pure metals. The first study related to the Mg-based amorphous alloys was conducted by Inoue et.al (Inoue et al., 1991). They accomplished to produce Mg-based amorphous alloys in a cylindrical form by casting melts into a copper mould. Wang et.al (Wang et al., 2012) reported the first time the MgZn-based amorphous alloys represent good functional ability compared to commercial crystalline Fe powders and crystalline MgZn with a greater corrosion resistance in water, in order to degrade azo dyes. They claimed that these alloys keep high reaction efficiency under even complex environmental conditions. Other studies related to Mg-based amorphous alloys are available in the literature (Iqbal and Wang, 2014; Luo et al., 2014; Zhao et al., 2014). The conducted some few studies of Mg-based amorphous alloys on textile azo dyes removal are compiled in Table 3. The degradation efficiency of azo dyes depends on pH of the reactive medium, the alloy dosages and initial dye concentration and dye volume (see Table 3).

*4.3 Some Miscellaneous-based Amorphous Alloys*

Synthesis of amorphous alloys needs strict circumstances including high vacuum, rapid cooling and high purity of raw materials for degrading azo dyes. It continues a need of developing tailored materials having high degradation efficiency, low cost and easy to fabrication. In this regard, some miscellaneous-based amorphous alloys such as Al, Co, Pd, Cu, Zn and Mn have been studied recently by engineers and scientist (Ben Mbarek et al., 2017; Das et al., 2016; Qin et al., 2015; Wang et al., 2017; Zhao et al., 2014). These metallic glasses are usually used in powder form that makes more active surface sites for highly efficient degradation of organic azo dye molecules when comparing to the crystalline metals of the same weight. According to the study conducted by Das et.al (Das et al., 2016), Alloying high percentage conducting aluminium with low percentage of transition metals as an amorphous form promotes catalytically active surface for the degradation of organic water pollutants without any toxic formation. Qin et.al (Qin et al., 2015) studied Co-based amorphous alloys produced by ball milling. Their study shows the studied alloys possess better resistance to corrosion in wastewater, little mass loss, durability and the highest reaction rate of all investigated powders. Mbarek et.al (Ben Mbarek et al., 2017) carried out the reduction reaction of the azo dye Reactive Black 5 by using Mn based amorphous alloys produced melt-spinning and ball-milling processes. In this system, the low activation energy and the rapid degradation kinetics was observed, which in turn makes promising as a low-cost, efficient material for azo dyes removal. Above studies are summarized in Table 4.

Removal efficiency of the various alloys is also shown as a function time in Fig. 7. According to Fig. 7, $Co_{78}Si_8B_{14}$ and $Mg_{65}Cu_{25}Y_{10}$ amorphous alloys have the most performance of removal efficiency at 2 min and 4 min, respectively. It takes more time for the other alloys because of the different production techniques and their compositions.

## 5. Conclusions and future perspectives

It is obvious that, freshwater is needed for living organism and if wastewater is treated properly, it can be used again. Azo dyes are main effluent of textile industry resulting huge volume of wastewater. Because of this reason effective treatment of azo dye effluents is needed. This review article presented amorphous alloys for azo dye removal from wastewater. Amorphous alloys have exhibited very promising results for the azo dye treatment from wastewater. Many researchers have been investigated azo dye degradation properties of some amorphous alloys with combination of different factors that are pH, temperature and dye-alloy dosage. The unique properties of amorphous alloys and their high efficient azo dye catalysis properties with their own and combined methods present a significant opportunity for dealing with azo dye effluents.

However, many challenges still exist in applicability. The future perspectives and suggestions are listed

- Most of the work has been carried out at laboratories scale. It is needed to adapting of these works in industry scale.
- Examination of the properties of amorphous alloys to degrade azo dyes is an important subject alone. However, the studies about ability of removal of hazardous contaminants such as heavy metals, nitrate arsenic, will be also needed by amorphous alloys.
- Reusability and production cost of amorphous alloys are needed to study more.
- Using more than one technique at azo dye degradation with metallic glasses is needed to investigate in terms of removal time, removal efficiency, large scale applicability and total required cost.

**Acknowledgements**

We would like to thanks Kahramanmaras Sutcu Imam University.

**Tables**

**Table 1**: Advantages and disadvantages of some waste dye removal techniques

| **Techniques** | **Advantages** | **Disadvantages** |
|---|---|---|
| Ion exchange | No absorbent loss | Not suitable for all dyes |
| Peat | Good adsorbent | Surface area is low |
| Fentons reagent | Effective decolourisation | Sludge generation |
| Absorption (Activated carbon) | High effective for various dyes | Very expensive |
| Ozonation | Effluent volume remains fix | Shorthalf life (20 min) |
| Membrane filtration | Removes all dyes | Concentrated sludge production |
| Photochemical | No sludge generation | Formation of byproducts |
| Irradiation | Effective oxidation at lab scale | Required a lot of dissolved $O_2$ |
| Electrokinetic coagulation | Economically applicable | High sludge generation |
| Biomass | Low operating cost and good efficiency | Slow process, performance depends on external factors |

**Table 2**: Azo dye degradation using Fe-Based Metallic Alloys

| **Alloy name** | **Production Methods** | **Dye** | **Integrated with other techniques** | **Initial concentrations** | **pH** | **Removal efficiency** | **Other details and Literatures** |
|---|---|---|---|---|---|---|---|
| $Fe_{78}Si_9B_{13}$ | Arc melting then roller spinning | Rhodamine B (RhB) | $H_2O_2$ | 20 mg/L | 3<br>3.5<br>4 | 70 % of RhB degrade within 60 min (at pH 3 and 0.5 M $H_2O_2$ addition | This alloy stable for at least four cycle and effect of pH, $H_2O_2$ and temperature were also studied, (Wang et al., 2014) |
| $Fe_{72}Si_2B_{20}Nb_6$ | Arc melting then roller spinning | Direct Blue 15 (DB 15) | $H_2O_2$ | 100 mg/L | | Fully transparent within 60 min (20 % hydrofluoric acid treatment for 40 min before reaction to obtain porous structure) | This study was also investigated, dye concentration, temperature, $H_2O_2$ concentration and different porous structures on azo dye degradation (Deng et al., 2017) |
| $Fe_{78}Si_{13}B_9$ | Amorphous alloy obtained from Antai Co. | Orange II<br><br>Methyl Orange<br><br>Direct Blue 6 | | 100<br>200 mg/L<br>300 | 2<br>4<br>10.3 | Fully transparent within 30 min (pH 2) also different pH, dye concentration and alloy concentration were studied. | Amorphous ribbon and same composition crystalline alloy and 300 mesh iron powders were examined (Tang et al., 2015) |
| $Fe_{73}Si_7B_{17}Nb_3$ | Arc melting then 1-high pressure argon gas atomization (GA) 2- Ball milling (BM) | Direct Blue 6 | | 200 mg/L | - | Half peak intensity for BM powders 10 min<br>GA powders 600 min<br>Fe powders 2000 min | Two different metallic glasses and 325 mesh iron powder were examined at various temperatures (Wang et al., 2012) |
| $Fe_{78}Si_9B_{13}$<br>$Fe_{73.5}Si_{13.5}B_9Cu_1Nb_3$ | Arc melting then roller spinning | Brilliant red 3B-A | $H_2O_2$<br>300W simulated solar light | 20 mg/L | 2<br>6.45<br>12 | $Fe_{78}Si_9B_{13}$ in 5 min<br>$Fe_{73.5}Si_{13.5}B_9Cu_1Nb_3$ in 20 min ( for 2g/L ribbon dose, 1.0 mM | Effect of pH, light intensity, catalyst dosage, $H_2O_2$ concentration were also studied (Jia et al., 2017) |
| | | | | | | $H_2O_2$ concentration and 7.7 μW/cm$^2$ irradiation intensity) | |
| $Fe_{78}Si_9B_{13}$<br>$Fe_{73.5}Si_{13.5}B_9Cu_1Nb_3$ | Vacuum melt spinning | Methyl blue<br>Methyl orange | $H_2O_2$<br>300W simulated solar light | | 2<br>3.4<br>7.8 | Full transparent $Fe_{78}Si_9B_{13}$ in 20 min at pH=2.<br>About 75 % transparent for $Fe_{73.5}Si_{13.5}B_9Cu_1Nb_3$ in 20 min at pH=2 | Effect of pH, $H_2O_2$ concentration, light intensity, dosage area and reusability were investigated (Jia et al., 2017) |
| $Fe_{78}(Si,B)_{22}$ | Melt spinning | Orange II | | 100 mg/L | | Full transparent within 60 min | Effect of temperature was also studied (Zhang et al., 2011) |
| $Fe_{78}Si_9B_{13}$ | Melt spinning | Cibacron brilliant red 3B-A (BR3B-A) | $H_2O_2$<br>300 W Xeon simulated solar light lamp | 10<br>20 mg/L<br>50<br>100 | 2<br>4<br>6.45<br>10 | Full transparent within 10 min (at 7.7 μW/cm$^2$ irradiation intensity, 20 mg/L dye concentration and pH=2 ) | Effect of pH, $H_2O_2$ concentration, light intensity, catalyst dosage and reusability were investigated (Jia et al., 2016) |
| $Fe_{78}Si_9B_{13}$ | Arc melting then roller spinning | Methylene blue | Solar light lamp and persulfate | 20 mg/L | 3.39 | Fully transparent within 20 min | Effect of persulfate, irradiation intensity, alloy dosage were studied. Outstanding sustainability of alloy was revealed (Jia et al., 2016). |
| A-$Fe_{84}B_{16}$<br>B-$Fe_{82}B_{18}$<br>C-$Fe_{80}B_{20}$ | Melt spinning | Direct Blue 6 | | 200 mg/L | 4<br>7<br>10.3 | Half life<br>A-6.30<br>B-6.13<br>C-4.44 | 300 mesh iron, annealed crystalline A alloy, temperature pH and alloy dosage were also mentioned (Tang et al., 2015) |
| $(Fe_{0.99}Mo_{0.01})_{78}Si_9B_{13}$ | Arc melting then roller spinning | Direct Blue 2B | | 200 mg/L | | Fully transparent within 30 min | Compare decolorization efficiency of annealed crystalline ribbon with amorphous ribbon and it was also examined reusability of ribbons (Zhang et al., 2010) |
| A-$Fe_{79}B_{16}Si_5$<br>B-$Fe_{66.3}B_{16.6}Y_{17.1}$ | Arc melting then roller spinning | Orange G | | 100 mg/L | 6 | A-Fully transparent within 180 min<br>B- About 50 % transparent within 180 min | Alloy B reusability up to 11$^{th}$ cycles whereas alloy A has 8 cycles reusability. In short, Alloy B was shorter decolorization time and more reusability compared to alloy B (Liu et al., 2014) |
| $Fe_{73.5}Si_{13.5}B_9Cu_1Nb_3$ | Melt spinning | Malachite green | 300 W simulated solar light lamp persulfate | 100 mg/L | 2.8 | Fully transparent within 30 min | Effect of pH, light intensity, catalyst dosage, persulfate Concentration and Stability and reusability of alloys were also studied (Liang et al., 2017) |
| A-$Fe_{76}B_{12}Si_9Y_3$ ribbon<br>B-prepared alloy at A ball milled<br>C-B annealed at 474 K for one hour<br>D- B annealed at 853 K for one hour<br>E-Crystalline zero-valent iron | Melt spinning then ball milling | Methyl orange | | 20 mg/L | 2<br>4<br>6<br>8<br>10<br>12 | Half-life for<br>A-90 min<br>B-5 min<br>C-10 min<br>D-20 min<br>E-5000 min | Effect of annealing temperature of alloys, pH, temperature and reusability were investigated. Also some results compared with zero valent iron (Xie et al., 2016) |
| $Fe_{78}Si_9B_{13}$ | Gas-atomized | Acid orange II | | 200 mg/L | 2.5<br>4.5<br>6.6<br>8.5<br>10.5 | 98 % transparent within 10 min | Effect of pH, reusability and degradation mechanism and kinetics were studied (Qin et al., 2017) |
| $Fe_{48}Cr_{15}Mo_{14}Y_2C_{15}B_6$ | Gas-atomized, ball milling | Direct Blue | | 15 mg/L | - | Transparent within 25 to 250 min according to milling time | (Das et al., 2015) |

**Table 3:** Azo dye degradation using Mg-Based Metallic Alloys

| **Alloy name** | **Production Methods** | **Dye** | **Initial concentrations** | **pH** | **Removal efficiency** | **Other details and Literatures** |
|---|---|---|---|---|---|---|
| $Mg_{60}Zn_{35}Ca_5$ | Ball milling | Congo red | 200 gm/L | 6.7 | Transparent after 15 min (with 0.2 g powders) | Effect of powder dosage was also investigated (Ramya et al., 2017) |
| $Mg_{65}Cu_{25}Y_{10}$ | Arc melting then melt spinning then ball milling | Direct Blue 6 | 20 mg/L | | Transparent after 4 min for Ball milled powder | Effect of nanoporous copper, ball milled powder and dealloyed powder were studied (Luo et al., 2014) |
| $Mg_{60}Zn_{35}Ca_5$ | Melt spinning | Direct Blue 6 | 200 mg/L | | Decolorization within 360 min | (Iqbal and Wang, 2014) |
| $Mg_{73}Zn_{21.5}Ca_{5.5}$ | Melt spinning | Direct Blue 6 | 200 mg/L | 3<br>7<br>10 | Transparent within 0.78 min | This study also examined three different particles annealed crystalline powder, power exposed air (oxygen) and effect of temperature and pH (Wang et al., 2012) |
| $Mg_{63+x}Zn_{32-x}Ca_5$ (x=0,3,7,10) | Gas-atomized and melt spinning | Direct Blue 6 | 200 mg/L | | M70-coded powder transparent within 7.15 min | The produced metallic glassy powders were considerably higher degradation capacity than Fe based powders found in this study. Also degradation rate were related to concentration of Mg gave best result at 70>73>66>63 respectively (Zhao et al., 2014). |

**Table 4:** Azo dye degradation using some miscellaneous alloys

| Alloy name | Production Methods | Dye | Initial concentrations | pH | Removal efficiency | Other details and Literatures |
|---|---|---|---|---|---|---|
| $Mn_{85}Al_{15}$ | Arc melting then roller spinning and then ball milling | Reactive Black 5 | 40 mg/L | 3<br>6<br>10 | Fully transparent within 30 min | Effect of temperature and pH were also studied (Ben Mbarek et al., 2017) |
| $Co_{78}Si_8B_{14}$ | Melt spinning then ball milling | - | 200 mg/L | 3<br>4<br>5<br>8<br>9<br>10 | Transparent less than 2 min | Four different powders, effect of pH, initial dye concentration, temperature and reusability were mentioned (Qin et al., 2015) |
| $Al_{91-x}Ni_9Y_x$<br><br>(x=0, 3, 6, 9) | Induction melting then melts spinning | Direct Blue 2B | 200 mg/L | 2<br>7<br>12 | Transparent less than 120 min | Effect of addition of Y to composition, temperature and pH were deeply investigated (Wang et al., 2017) |
| $Al_{82}Y_8Ni_7Fe_3$ | Gas-atomized | Direct Blue 6 | 0.015 Molar | | transparent less than 40 min | The produced alloy was compared with zero valent iron and show superior properties compared to it (Das et al., 2016) |
| A-AlCoCrTiZn<br><br>B- AlCoCrFeNi<br><br>C- CoCrFeMnNi | Ball milling | Direct Blue 6<br>Orange II | 200 mg/L | | Transparent less than 10 min for alloy A | Alloy A has been one of the best metallic glass in azo dye degradation area. This was also the only study with High Enthalpy Alloys and azo dye degradation (Lv et al., 2016) |

**Table Captions**

Table 1: Advantages and disadvantages of some waste dye removal techniques

Table 2: Azo dye degradation using Fe-Based Metallic Alloys

Table 3: Azo dye degradation using Mg-Based Metallic Alloys

Table 4: Azo dye degradation using some miscellaneous alloys

**Figures Captions**

Figure 1: Distribution of total water on world, freshwater and human accessible freshwater.

Figure 2: Number of publications with “wastewater” mentioned in the title.

Figure 3: Schematic diagram of degradation performance of azo dyes for amorphous alloys.

Figure 4: Schematic diagram of melt spinning technique.

Figure 5: Schematic diagram of mechanical alloying technique.

Figure 6: Schematic diagram of furnace in arc melting machine for suction casting technique.

Figure 7: Removal efficiency of the different alloys as a function time.

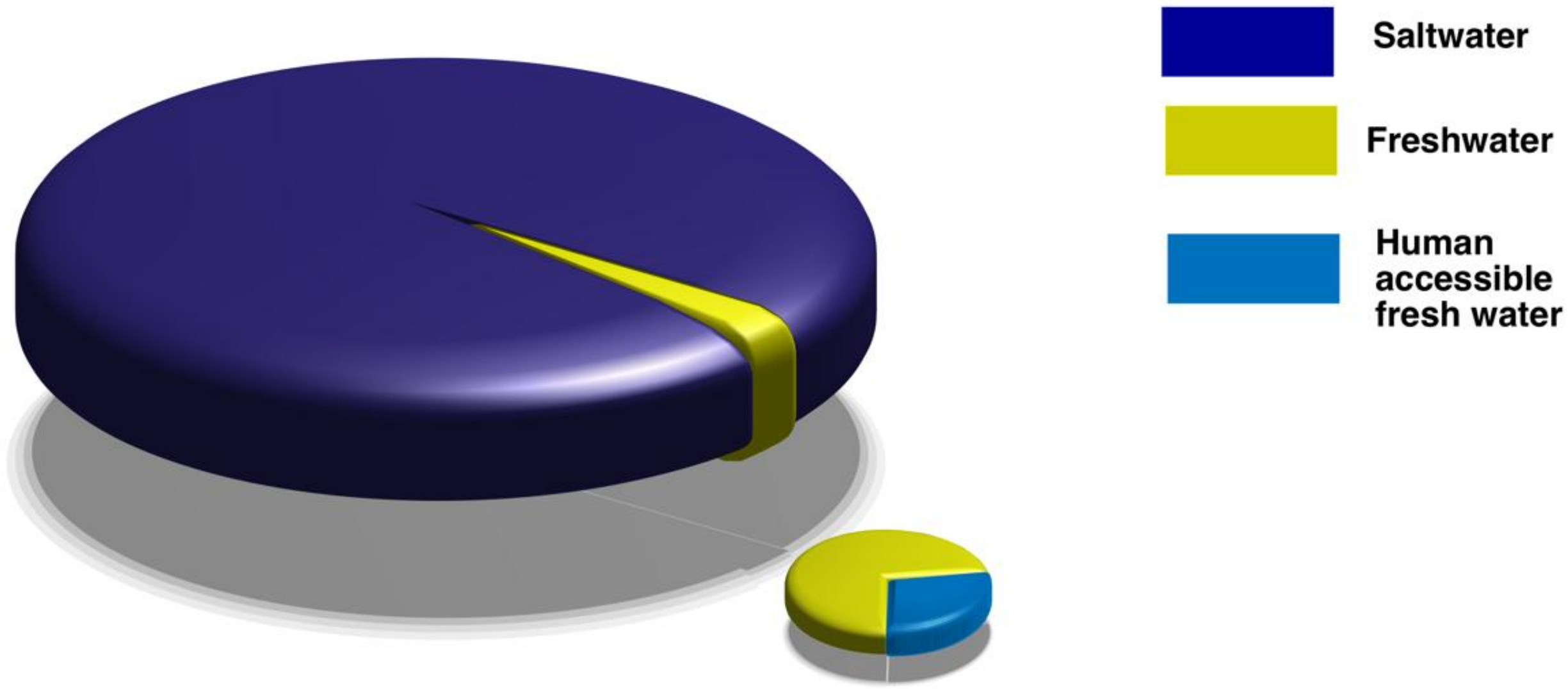

Figure 1

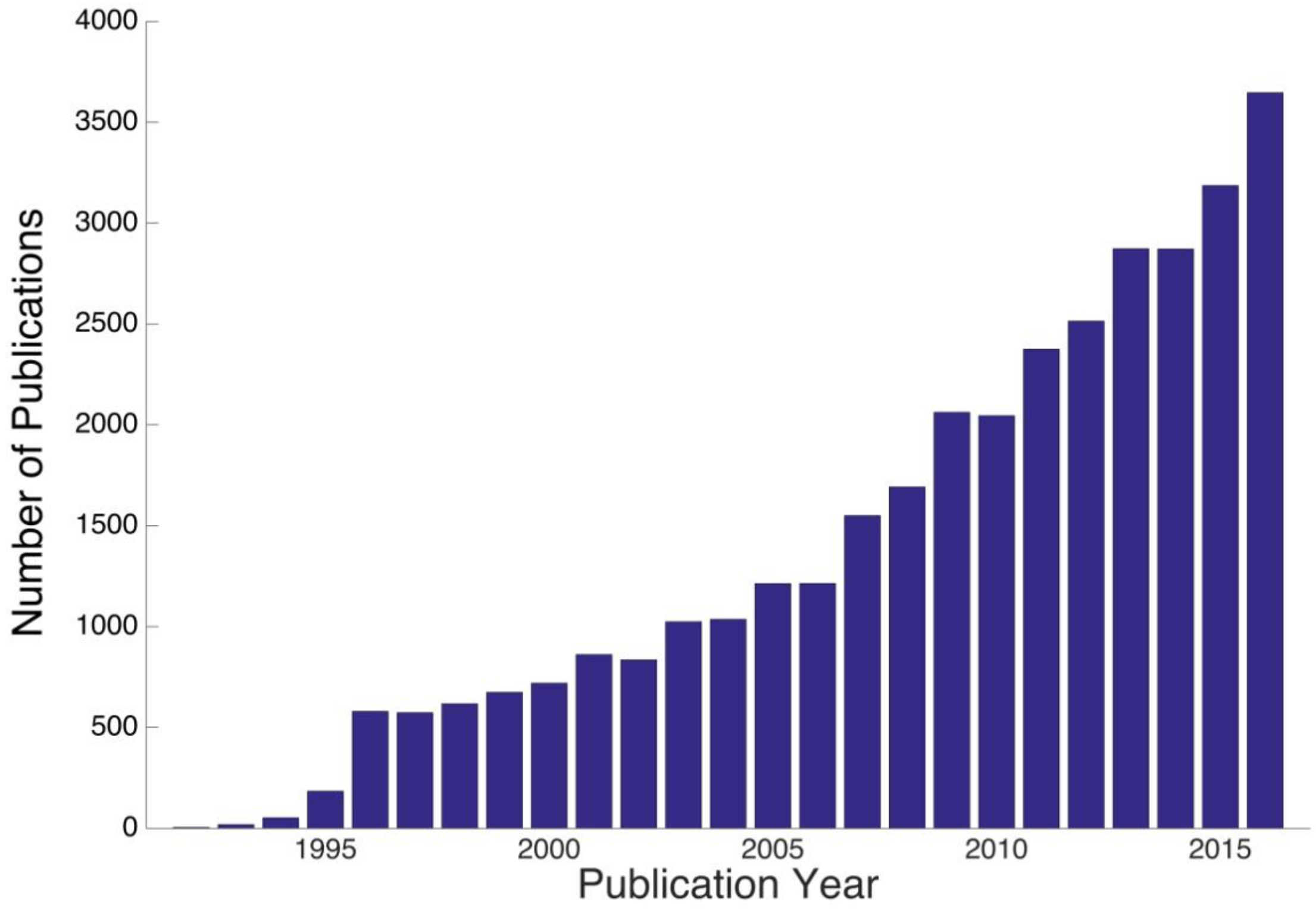

Figure 2

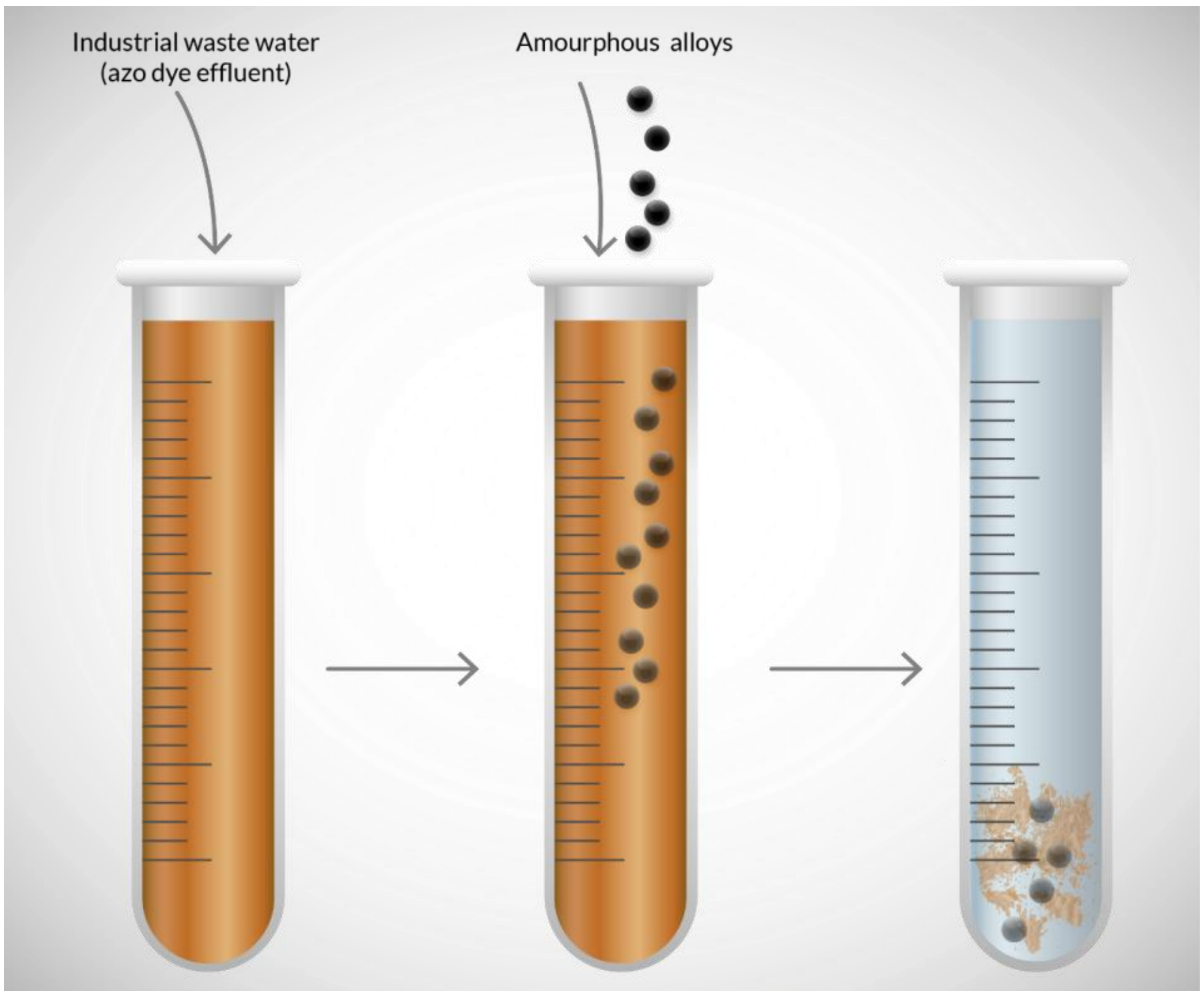
Industrial waste water
(azo dye effluent)
Amourphous alloys

Figure 3

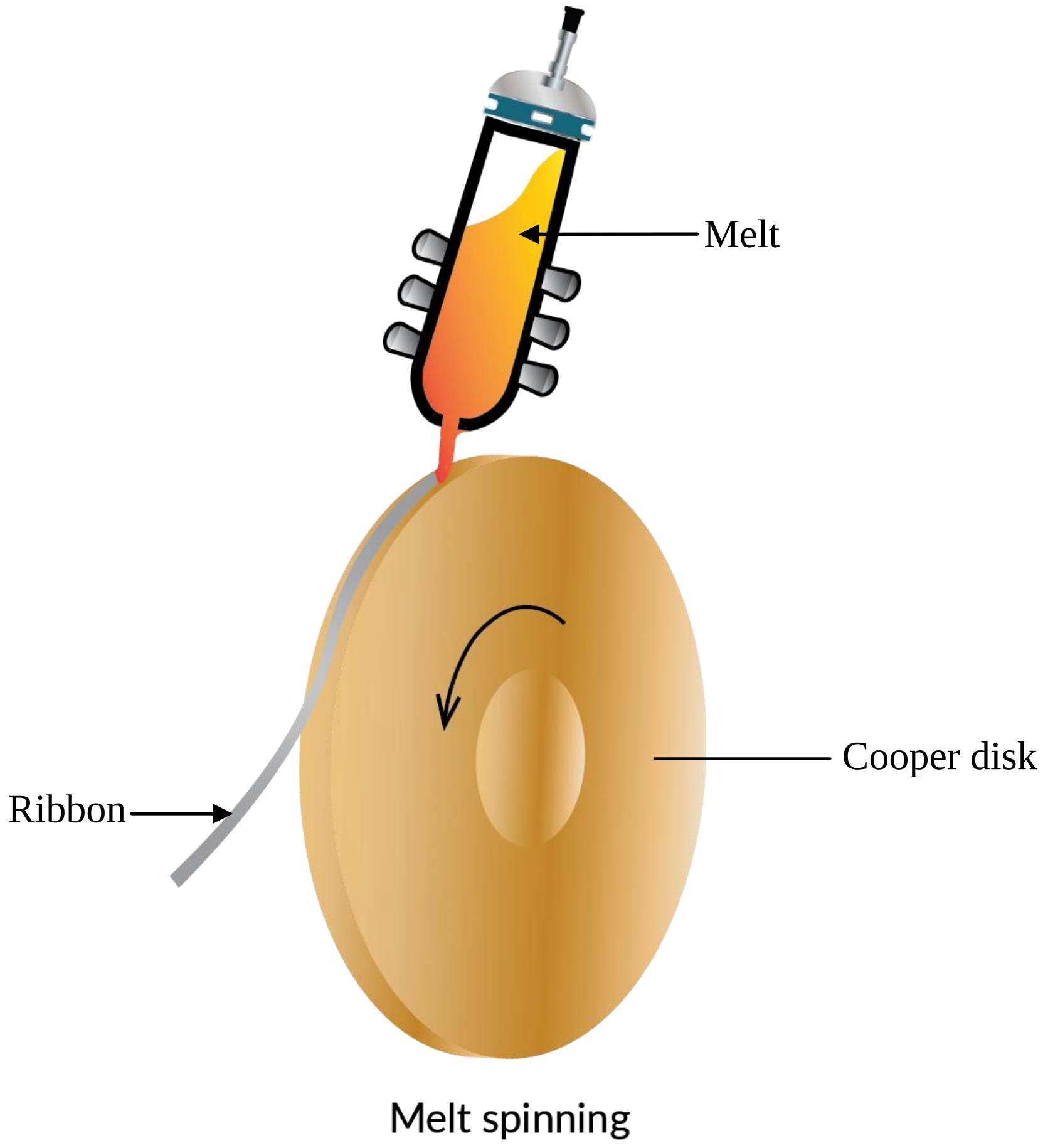

Figure 4

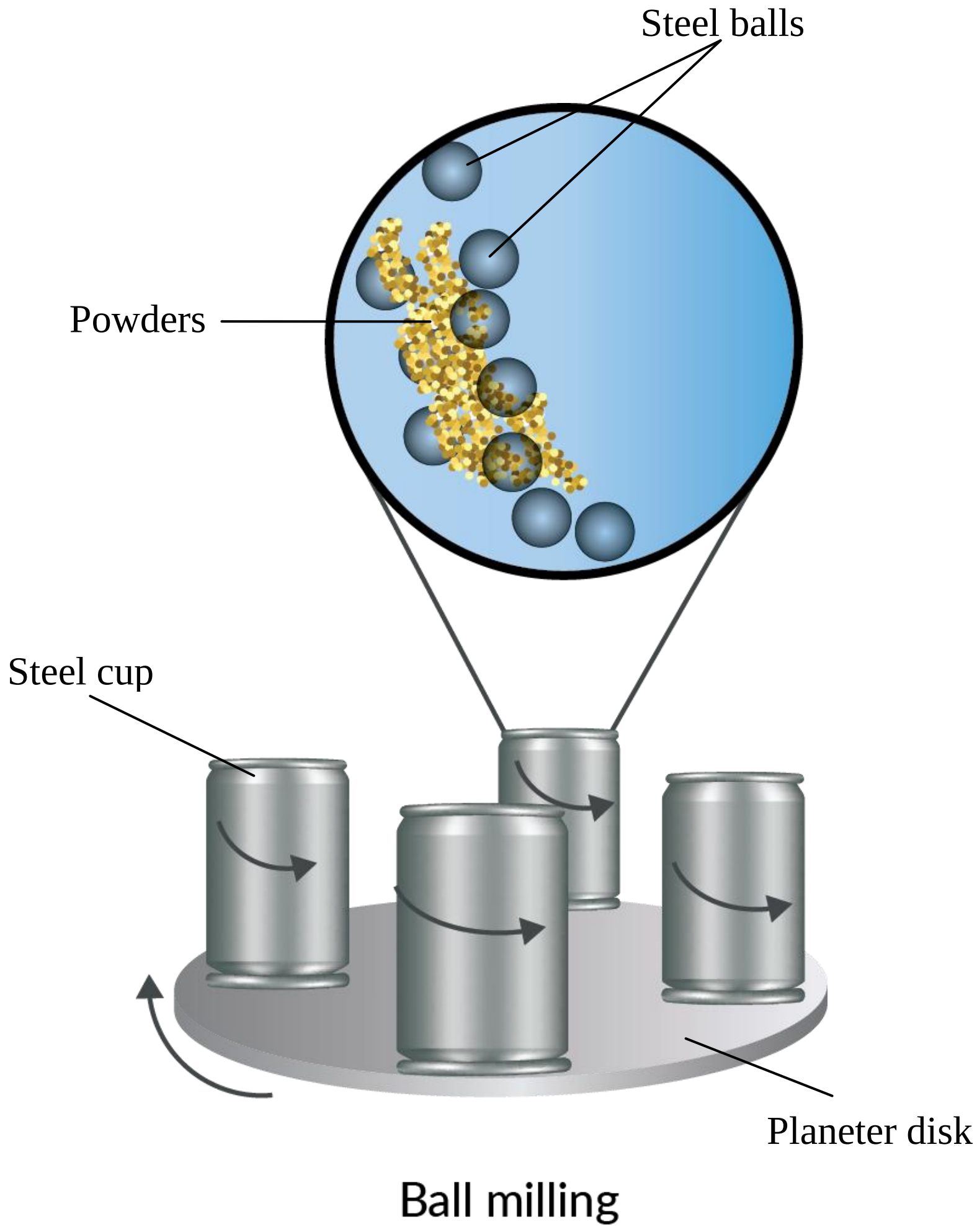

Figure 5

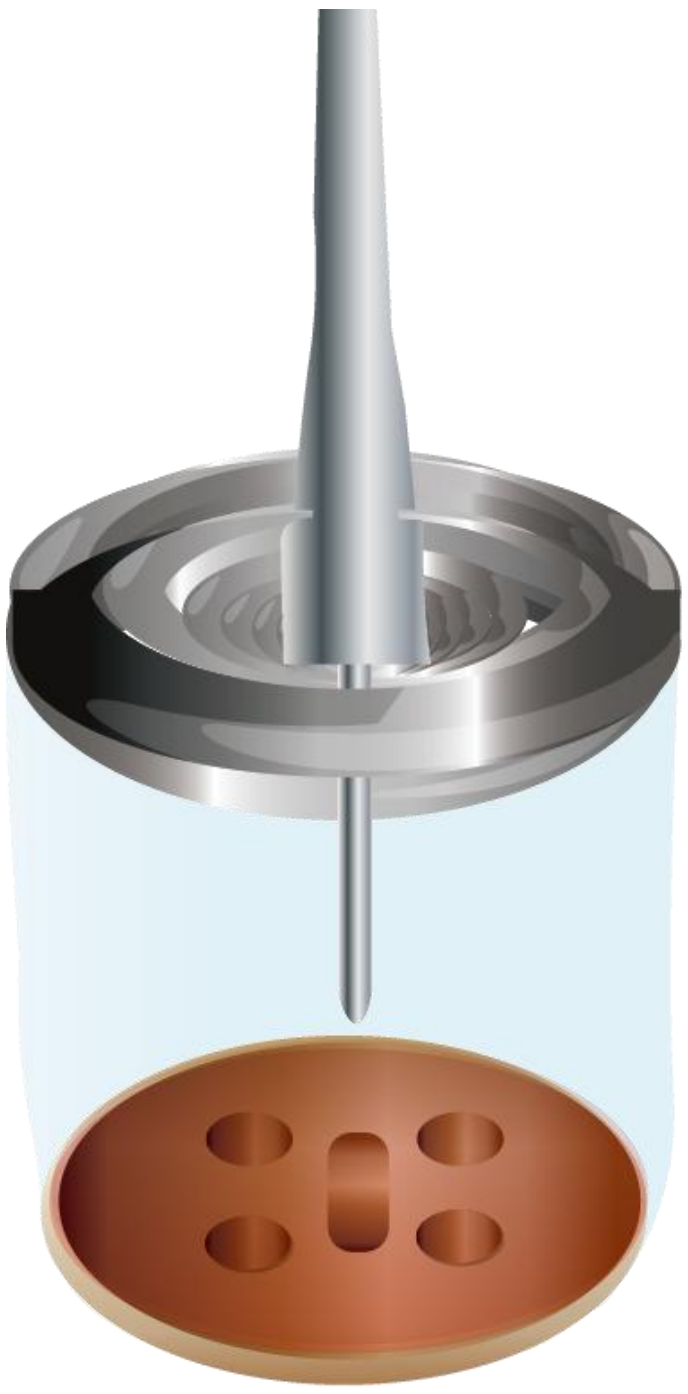

Arc melting furnace

Figure 6

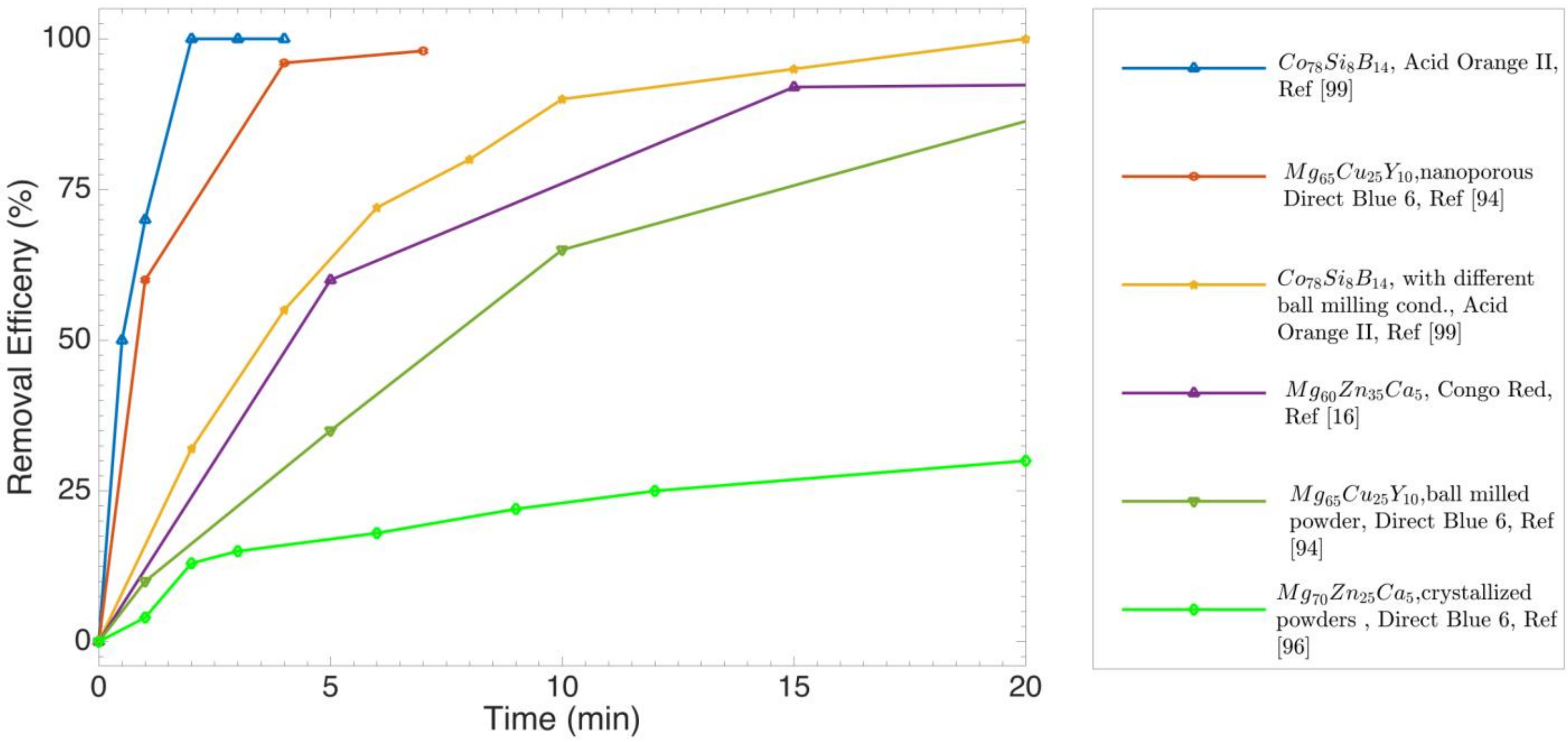

Removal Efficeny (%)
Time (min)
0
25
50
75
100
0
5
10
15
20
$Co_{78}Si_8B_{14}$, Acid Orange II, Ref [99]
$Mg_{65}Cu_{25}Y_{10}$,nanoporous Direct Blue 6, Ref [94]
$Co_{78}Si_8B_{14}$, with different ball milling cond., Acid Orange II, Ref [99]
$Mg_{60}Zn_{35}Ca_5$, Congo Red, Ref [16]
$Mg_{65}Cu_{25}Y_{10}$,ball milled powder, Direct Blue 6, Ref [94]
$Mg_{70}Zn_{25}Ca_5$,crystallized powders , Direct Blue 6, Ref [96]

Figure 7